\documentstyle[12pt]{article}
\newcommand{\nc}{\newcommand}
\nc{\beq}{\begin{equation}} \nc{\eeq}{\end{equation}}
\nc{\beqa}{\begin{eqnarray}} \nc{\eeqa}{\end{eqnarray}}
\nc{\lsim}{\begin{array}{c}\,\sim\vspace{-21pt}\\< \end{array}}
\nc{\gsim}{\begin{array}{c}\sim\vspace{-21pt}\\> \end{array}}

\nc{\scR}{{\cal R}}
\nc{\scL}{{\cal L}}
\nc{\al}{\alpha}
\nc{\ald}{\dot{\alpha}}
\nc{\be}{\beta}
\nc{\bed}{\dot{\beta}}
\nc{\lam}{\lambda}
\nc{\nud}{\dot{\nu}}
\nc{\lamd}{\dot{\lam}}

\newcommand{\drawsquare}[2]{\hbox{%
\rule{#2pt}{#1pt}\hskip-#2pt
\rule{#1pt}{#2pt}\hskip-#1pt
\rule[#1pt]{#1pt}{#2pt}}\rule[#1pt]{#2pt}{#2pt}\hskip-#2pt
\rule{#2pt}{#1pt}}

\newcommand{\Yfund}{\raisebox{-.5pt}{\drawsquare{6.5}{0.4}}}

\nc{\Ap}[2]{A^\prime_{#1#2}}

\nc{\Q}[2]{Q_{#1#2}}
\nc{\R}[2]{R_{#1#2}}
\nc{\Y}[2]{Y_{#1#2}}
\nc{\V}[2]{V_{#1#2}}
\nc{\q}[2]{q_{#1#2}}
\nc{\G}[2]{G_{#1#2}}
\nc{\W}[2]{W_{#1#2}}
\nc{\D}[2]{D_{#1#2}}
\nc{\A}[2]{A_{#1#2}}
\nc{\p}[2]{p_{#1#2}}
\nc{\vv}[2]{v_{#1}^{#2}}
\nc{\rr}[2]{r_{#1}^{#2}}
\nc{\lij}[2]{l_{#1}^{#2}}

\nc{\spa}{SU(2)_1}
\nc{\spb}{SU(2)_2}
\nc{\spc}{SP(2n-4)}
\nc{\spd}{SP(4n+2m-10)}

\nc{\Lsc}[2]{\scL_{#1#2}}
\nc{\Lrm}[2]{L_{#1#2}}
\nc{\Rsc}[2]{\scR_{#1#2}}

\begin{document}

\begin{titlepage}

\begin{center}

\vspace{2cm}

{\hbox to\hsize{hep-th/9606184 \hfill EFI-96-24}}
{\hbox to\hsize{\hfill Fermilab-Pub-96/157-T}}
{\hbox to\hsize{\hfill revised version }}
\bigskip

\vspace{2cm}

\bigskip

\bigskip

\bigskip

{\Large \bf  Supersymmetry Breaking and Duality in  
${\bf SU(N) \times SU(N - M)}$ Theories}

\bigskip

\bigskip

{\bf Erich Poppitz}$^{\bf a}$, {\bf Yael Shadmi}$^{\bf b, c}$
and  {\bf Sandip P. Trivedi}$^{\bf b, d}$ \\

\bigskip

\bigskip

$^{\bf a}${\small \it Enrico Fermi Institute\\
 University of Chicago\\
 5640 S. Ellis Avenue\\
 Chicago, IL 60637, USA\\

{\rm email}: epoppitz@yukawa.uchicago.edu\\}
\smallskip

 \bigskip

$^{\bf b}${ \small \it Fermi National Accelerator Laboratory\\
  P.O.Box 500, Batavia\\
  IL 60510, USA\\

$^{\bf c}${\rm email}: yael@fnth06.fnal.gov \\

$^{\bf d}${\rm email}: trivedi@fnal.gov\\ }

\vspace{1.3cm}

{\bf Abstract}

\end{center}

We consider a  class of $N=1$ supersymmetric Yang-Mills theories,
with gauge group $SU(N) \times SU(N - M)$ and fundamental
matter content.
Duality plays an essential role  in analyzing the nonperturbative infrared
dynamics of these models.  We find that Yukawa couplings drive these 
theories into the confining phase, and show how the nonperturbative
superpotentials arise in the dual picture. We show that the odd-$N$,
 $M = 2$ models with an appropriate tree-level superpotential 
break supersymmetry.

\end{titlepage}

\renewcommand{\thepage}{\arabic{page}}
\setcounter{page}{1}

\baselineskip=18pt

\section{Introduction. }

In order to be relevant to nature, supersymmetry must be spontaneously
broken.  One  attractive idea is that the breaking occurs
nonperturbatively.  The  electroweak  scale could then arise from
the Planck scale through the logarithmic running of coupling
constants and this would help explain its smallness \cite{witten}.

The past few years have seen  remarkable progress in the study
 of the non-perturbative behavior of SUSY gauge theories \cite{seibergexact},
\cite{seiberg}, \cite{dualityreview}.
This progress has led in turn to a better understanding of
nonperturbative SUSY breaking \cite{ads}. In particular,  many 
new examples of
theories exhibiting this phenomenon have  been found \cite{iss}--\cite{US}.

In this paper we will study a simple class of  $SU(N) \times SU(N - M)$
gauge theories with matter in  the fundamental representation,
and show that they break supersymmetry dynamically.
These models are a generalization of the $SU(N) \times SU(N-1) $ theories
discussed in an earlier paper \cite{US}.
Various  ideas developed  in that paper will prove useful for 
studying the $SU(N) \times  SU(N-M)$ theories.
In particular,  unlike the models considered previously,
we find that  product-group duality will be
essential  in  understanding the non-perturbative behavior of
the theories with 
$M > 1$\footnote{Duality in the context
 of supersymmetry
breaking has been discussed in \cite{scott}.}. For example, duality will help
identify the  appropriate degrees of freedom
in the low-energy theory, in terms of which the K\" ahler potential is
nonsingular.
We will also see, as noted in \cite{lisa},  \cite{US},
that  Yukawa couplings will  sometimes  
drive these theories into  the confining regime, and that this  
will play
a crucial role in the breaking of supersymmetry.

This paper is organized as follows.  First we introduce the 
$SU(N) \times SU(N -M)$ models.  
We then  discuss their
$SU(N) \times SU(M)$ duals in Sect. 3. 
In Sect. 4 we show how the nonperturbative superpotential  
arises in the dual picture.  In Sect. 5, we  turn to the question of 
supersymmetry  breaking. In particular, we show  that  the  $M=2$,
 odd-$N$ models break supersymmetry, once  appropriate Yukawa couplings are
added. 
 Finally,  in Sect. 6,   
we discuss the supersymmetry preserving  $M=0$ models.

\section{The ${\bf SU(N) \times SU(N-M)}$ Models.}

The theories we consider in this paper have an $SU(N) \times SU(N-M)$
gauge symmetry with matter content consisting of a single field,
$Q_{\alpha {\dot{\alpha}}}$, that transforms as (\Yfund , \Yfund ) under the
gauge groups, $N-M$ fields, $\bar{L}^\alpha_I$, transforming as
$(\overline{\Yfund}, {\bf 1})$, and $N$ fields,  $\bar{R}^{\ald}_A$,
that  transform as  $({\bf 1}, \overline{\Yfund})$.  Here, as in the
subsequent discussion,  we denote the gauge indices of  $SU(N)$
and $SU(N-M)$  by $\alpha$ and $\ald$, respectively, while
$I = 1\ldots N-M$ and $A = 1 \ldots N$ are  flavor indices. We note that
these theories  are  chiral---no mass terms can be added for any of the matter
fields. The models with $M=1$ were considered in \cite{US}; here we study
the $M > 1$ models.

We  begin our study by considering the classical moduli
space. It is described by the gauge invariant chiral superfields
$Y_{IA} = \bar{L}_I \cdot Q \cdot \bar{R}_A$,
$\bar{b}^{A_1 \ldots A_M} = (\bar{R}^{N - M})^{ A_1 \ldots A_M }$
and
$\bar{\cal{B}} = Q^{N - M} \cdot \bar{L}^{N - M}$ (when appropriate,
all indices are contracted with  $\epsilon$-tensors),
subject to the classical constraints $Y_{I A_1} \bar{b}^{A_1 \ldots A_M} = 0$
and
$\bar{b}^{A_1 \ldots A_M}  \bar{\cal{B}}  \sim (Y^{N-M})^{A_1 \ldots A_M}$.
It is easy to see \cite{lutywati}, \cite{US},
that the classical  superpotential
\beq
\label{wtree}
W_{tree} = \lambda^{IA} ~ Y_{IA}
\eeq
with  maximal rank Yukawa-coupling matrix
lifts all classical flat directions with the  exception of the $SU(N-M)$
baryons $\bar{b}^{A_1 \ldots A_M}$.
Along the classical $\bar{b}^{A_1 \ldots A_M} \ne 0 $ flat direction the
$SU(N)$ gauge group is
completely unbroken and one expects strong quantum effects to be
important. 

The baryonic flat direction can be lifted, for $M=2$, by adding the 
tree-level superpotential 
\beq
\label{wtreebaryon}
W_{tree} = \lambda^{IA} ~ Y_{IA} + \alpha_{AB} ~\bar{b}^{AB} ~.
\eeq
As above $\lambda_{IA}$ has to have maximal rank, i.e. in this case rank $N-2$. 
The matrix $\alpha_{AB}$ needs to satisy two conditions. First it has 
to be invertible. Second, the projection of $\alpha_{AB}$ into the 
cokernel of $\lambda^{IA}$ needs to be invertible as well. The last 
condition can be stated more explicitly as follows. 
By flavor rotations one can go to a basis in which $\lambda_{IA}= ~ \lambda_I
\delta_{IA} $ for $A<N-1$ and $\lambda_{I(N-1)}$ and $\lambda_{IN} =0$. 
In this basis, the $N-2$ dimensional  matrix formed from 
$\alpha_{AB}$ by restricting 
$A$ and $B$ to be  $\le N-2$ must be invertible. 
For even $N$, there is no 
nonanomalous $R$ symmetry
which is  left unbroken by the superpotential (\ref{wtreebaryon}). As we will see
 below, the 
even-$N$, $M = 2$ models do not break supersymmetry, in conformity with the 
criteria of ref. \cite{nelsonandseiberg}.

For $N$-odd, $M = 2$, the matrix $\alpha_{AB}$ has to be of maximal 
rank ($N-1$), 
and  its cokernel should
contain the cokernel of $\lambda^{IA}$ (rank $\lambda = N - 2$). 
As opposed to the even-$N$ case, the superpotential (\ref{wtreebaryon})
that lifts all flat directions preserves a nonanomalous,
flavor dependent,  $R$ symmetry. To see that,  choose
for example $\alpha^{AN} = 0, \lambda^{I N} = \lambda^{I (N-1)} = 0$ 
(to lift the classical flat 
directions). Then 
one sees that
the field $\bar{R}_N$ appears in each of the baryonic terms of the 
superpotential
 (\ref{wtreebaryon}), while it does not appear in any of the Yukawa terms.
Assigning different $R$ charges to the four types of fields, $\bar{R}_N$,
$\bar{R}_{A < N}$, $Q$, and $\bar{L}_I$, one has to satisfy four conditions:
two conditions ensuring that the superpotential (\ref{wtreebaryon}) 
has $R$ charge 2, and
two conditions that the gauge anomalies of this $R$ symmetry vanish. 
It is easy to see that
there is a unique solution to these four conditions (a similar 
flavor-dependent $R$ symmetry
is preserved in the $M =1 $ models, for all $N$, when all classical flat 
directions are lifted
\cite{US}).

Lifting the baryonic flat directions for $M>2$ is much more complicated and 
we have not been able to analyze this fully yet.

\section{The ${\bf SU(N) \times SU(M)}$ Dual.}

We begin our analysis of the quantum theory by noting that
it has a dual description in terms of an $SU(N)\times SU(M)$ gauge
theory.
This dual theory may be constructed as follows: first, turn off the
$SU(N)$ coupling. The electric theory is then $SU(N-M)$ with $N$ flavors,
whose dual is an $SU(M)$ gauge theory.
Turning the $SU(N)$ coupling back on,
the $SU(N)\times SU(M)$ dual is obtained. 
We will find it useful to study the low-energy dynamics by 
analyzing this dual theory.
Duality will help us identify the
low-energy degrees of freedom in terms of which the K\" ahler potential
is nonsingular.
%
%
\begin{table} \begin{center}
{\centerline{Table 1: The field content of the dual
$SU(N) \times SU(M)$ theory.}}
\label{table33}
\begin{tabular}{c|c|c} \hline \hline
$\ $ &$SU(N) $& $SU(M) $ \\
\hline
$ q^\alpha_\nu $ & $\overline{\Yfund} $& \Yfund  \\
$\bar{r}^{A \nu}$ &{\bf 1} & $\overline{\Yfund}$ \\
${1 \over \mu} M_{\alpha A} $ & \Yfund & {\bf 1}  \\
$\bar{L}^{\alpha}_I$& $\overline{\Yfund}$& {\bf 1}  \\
\hline
\end{tabular}
\end{center}
\end{table}
%

Before doing so, however,  two comments
regarding the various length scales  present
are in order.
First,
while the dual theory was constructed in the limit
 $\Lambda_2 \gg \Lambda_1$, prior experience,
based on a study of an $SU(2) \times SU(2)$
theory~\cite{US},
strongly suggests that  the two  theories
are equivalent in the
infra-red for
all  values of the ratio $\Lambda_2/\Lambda_1$.
Therefore,
the infra-red behavior of the electric theory can be understood  by
studying the dual theory.

Second, in our discussion we
will assume  that the SUSY breaking scale is much lower than
the strong coupling scales of both groups.
Our analysis will show   that
SUSY breaking occurs only in the presence of the Yukawa couplings
eq.~(\ref{wtreebaryon}).  Thus, by making the couplings $\lambda^{IA}$  small
enough, one expects that  the SUSY breaking scale can be
made arbitrarily small too,
and  in this regime  our  analysis   will be  self-consistent.

We now go on  to consider the dual theory in some detail.
The field content of the dual theory is summarized in Table 1.
We see that the fields  $q^{\alpha}_{\nu}$ and $\bar{r}^{A \nu}$ are dual
to $Q_{\alpha \ald}$ and $ \bar{R}_A^{\ald}$ respectively, while the
$SU(M)$ singlet,  $M_{\alpha A}$, is dual to the $SU(N-M)$ meson,
$M_{\alpha A}$. Here $\nu = 1 \ldots M$ is
the $SU(M)$  index.
The dual theory has a Yukawa superpotential \cite{seiberg}:
\beq
\label{wdual1}
W = {1 \over \mu} ~ M_{\alpha A}~ \bar{r}^A \cdot q^{\alpha}~.
\eeq
The dimension-one parameter $\mu$ and
the strong coupling scales $\Lambda_2$ and $\bar{\Lambda}_2$ 
of the electric and
 magnetic
theories obey the matching 
relation\footnote{Hereafter we leave out the numerical constants appearing
in the various scale matching relations. These constants are calculable but 
are not essential to the present discussion.} 
\cite{seiberg}: 
\beq
\label{sumscalematching}
\Lambda_2^{2 N - 3 M} \bar{\Lambda}_2^{ 3 M - N} \sim \mu^{N}~.
\eeq
Symmetries also allow us to relate the scale of the $SU(N)$ group
in the dual theory,
$\bar{\Lambda}_1$,
to the scales of  the electric theory:
\beq
\label{sunscalematching}
\bar{\Lambda}_{1}^{2 N} \sim
\Lambda_1^{2 N + M} ~ { \Lambda_2^{2 N -3 M \over 2}
\over \mu^{ N - {M\over 2}}}\ .
\eeq

Before proceeding, one comment about the dual theory  is worth  making.
Note that we started with an electric theory that was  chiral  but in contrast the 
dual theory is  {\it not } chiral. From  Table 1
 we see that   mass terms  can be added for the $\bar{L}_I^{\alpha}$ and the
$M_{\alpha  A} $ fields.  As we will see
below,  these mass terms  will correspond to Yukawa couplings in
the electric theory. Examples of duality between chiral and non-chiral theories have
also  been found  in   \cite{pouliotanti} and \cite{pouliotstrassler}. 

 We now proceed with the discussion of the dual theory.  
 In this theory $SU(N)$ has $N$ flavors (see
Table 1) and is
 therefore confining in the infra-red. At scales below
$\bar{\Lambda}_1$,
the light degrees of freedom are the $SU(N)$
mesons:
\beqa
\label{sameson}
N_{A \nu} &=& {1\over\mu}\, M_{\alpha A} q^\alpha_\nu  
\ , ~\nonumber\\
K_{A I} &=&
{1\over\mu}\, M_{\alpha A} \bar{L}^\alpha_I
= {1\over\mu}\, Y_{IA} \ ,
~\nonumber
\eeqa
and the $SU(N)$ baryons:
\beqa
\label{baryonmap}
\cal{B} &=& {\rm det} (M_{\alpha A}/\mu)~, \nonumber \\
\bar{\cal{B}}^\prime &=& q^{M} \cdot \bar{L}^{N-M} \\
& \sim & \mu^{M\over 2} ~\Lambda_2^{3 M - 2 N \over 2}
~Q^{N - M} \cdot \bar{L}^{N - M} =
 \mu^{M \over 2} ~\Lambda_2^{3 M - 2N \over 2} ~\bar{\cal{B}}
~\nonumber.
\eeqa
In the last equation we  used  the baryon operator duality map
of SQCD \cite{seiberg}, \cite{dualityreview}.
Note that the field $\cal{B}$ in (\ref{baryonmap})
vanishes classically in the electric  theory.  We note also that  the 
second equation in (\ref{sameson})  relates 
${1\over\mu}\, M_{\alpha A} \bar{L}^\alpha_I$,  which is a mass term 
in the dual theory,  to  $Y_{IA}$, which is a Yukawa coupling
in the electric theory. 

Adding the confining superpotential of the $SU(N)$ theory
to~(\ref{wdual1}), the total superpotential,
written in terms of the $SU(N)$ mesons and baryons, becomes
\beq
\label{wdual2}
W = ~ N_{A \nu} \bar{r}^{A \nu} +
{\cal {A} }\left( N^M \cdot K^{N - M} - {\cal{B}} ~
\bar{\cal{B}}^\prime - \Lambda_{1 L}^
{2 N} \right) ~,
\eeq
with ${\cal {A} }$ a Lagrange multiplier superfield.

Below the scale  $\bar{\Lambda}_1$,
the  $SU(M)$ theory has $N$ flavors: 
$N_{A }/\bar{\Lambda}_{1}$
and $\bar{r}^{A}$.
It  therefore seems, at first glance, to be in the dual regime.
However,  the confining dynamics of the $SU(N)$ theory has
turned the Yukawa couplings of eq.~(\ref{wdual1}) into
mass terms for  the fields $N_{A }/\bar{\Lambda}_{1}$,
and  $\bar{r}^{A}$, see  eq.~(\ref{wdual2})\footnote{On accounting for the 
correct normalization of the field
$N_{A}/\bar{\Lambda}_{1}$, its mass term is seen to be of order
$\bar{\Lambda}_1$ .}.
Thus, the $SU(M)$ theory  too is driven into the confining regime.
This will play an important role in our discussion of SUSY breaking below.
One feature of this low-energy $SU(M)$ theory is that its
scale $\bar{\Lambda}_{2L}$ is  field dependent. We determine 
this scale in the next section and show how it helps recover, in the dual 
theory, some features of the electric theory.

\section{Gaugino Condensation in the Dual Picture.}

Below the confining scale of $SU(N)$, we have an $SU(M)$ theory
with $N$ flavors.
The scale of this theory, which we denote by
$\bar{\Lambda}_{2 L}$,
can be determined by symmetry considerations  to be
\beq
\label{lambda2l}
\bar{\Lambda}_{2 L}^{3 M - N} \sim \bar{\Lambda}_2^{3 M - N} ~
{ {\cal{B}} \over  \bar{\Lambda}_{1}^{N} }
~f \left( { \bar{\Lambda}_1^{2N}
\over {\cal{B}}~ \bar{\cal{B}}^\prime } \right) ~,
\eeq
where  $f$  is an unknown function which will  be determined below.
The physics behind this expression was explained in ref.~\cite{US}: the field
dependence of the
scale of the low-energy $SU(M)$ theory is required by Green-Schwarz
anomaly cancellation.
A $U(1)$ symmetry, which is anomaly free
in the ultraviolet, above the $SU(N)$ confinement scale, has an $SU(M)$ anomaly
in the infrared, below the $SU(N)$ confinement scale.
The anomaly is cancelled by the shift of an
axion-dilaton superfield, proportional to ${\cal{B}} f$ in
eq.~(\ref{lambda2l}).
 
The dilaton superfield can be determined using duality.
Since an analogous calculation is described in
Section~3.2 of ref.~\cite{US}, we only outline the main idea here.
While the dilaton superfield of eq.~(\ref{lambda2l}) is generated by
strong-coupling (confining) dynamics,
one can construct a dual theory in which the dilaton arises
as a weak-coupling (``higgsing'') effect.
Specifically, adding one extra flavor of $SU(N)$ to the $SU(N)\times SU(M)$
theory discussed above,
one can construct a dual of this theory with gauge group
 $SU(N)\times SU(N-M+1)$. In this theory  $SU(N)$ again confines. 
But this time, below its
confining scale, one is left with an $SU(N-M+1)$ gauge theory whose scale
is field-{\it independent}. On giving  a mass to the extra $SU(N)$ flavor
we added,  $SU(N-M+1)$ is higgsed to $SU(N-M)$.
The relevant vevs, and therefore the $SU(N-M)$ scale, depend on the $SU(N)$
baryon  ${\cal{B}}$.
But this $SU(N-M)$ is precisely the dual of the $SU(M)$ with which
we started, so that its scale determines the scale of
$SU(M)$,   $\bar{\Lambda}_{2 L}$.
On carrying out this exercise, one finds that
$f \equiv 1$
in  eq.~(\ref{lambda2l}),   giving the relation:
 \beq
\label{lambda2S}
\bar{\Lambda}_{2 L}^{3 M - N} \sim \bar{\Lambda}_2^{3 M - N} ~
{ {\cal{B}} \over  {\bar \Lambda}_1^{N} } \ .
\eeq

As was mentioned in the previous section, the fields $N_{A }/\bar{\Lambda}_1$ 
and $\bar{r}^{A}$ acquire mass. In order to obtain the superpotential after 
they are integrated out one can proceed, somewhat heuristically, as follows. 
Since all $N$ flavors are heavy we expect to be left with
a pure $SU(M)$  Yang-Mills  theory at low energies.  The scale of  this
theory can be determined  by the standard SQCD scale matching relation,
at the scale of the heavy quark mass $\bar{\Lambda}_1$,  to be:
\beq
\label{lambda2ll}
\bar{\Lambda}_{2 LL}^{3 M} \sim \bar{\Lambda}_{2 L}^{3 M - N}
\bar{\Lambda}_1^N.
\eeq
Gaugino condensation in the pure $SU(M)$ theory generates
a superpotential \cite{seibergexact}:
\beq
\label{wdual3}
W_{LL} \sim  \bar{\Lambda}_{2 LL}^3 \sim
\left( \bar{\Lambda}_2^{3M-N} ~ {\cal{B}}  \right)^{1 \over M} ,
\eeq
where in the last expression we have substituted for $\bar{\Lambda}_{2 LL}$
from eq. (\ref{lambda2S})    and eq. (\ref{lambda2ll}).
One then expects the full superpotential to be given by adding the terms
that remain in eq. (\ref{wdual2})  to the
superpotential induced by gaugino condensation, eq. (\ref{wdual3}).

The fields that remain in the theory after integrating out the
heavy quarks are the $SU(N)$ mesons $K_{IA} \sim Y_{IA}$,  the
antibaryon
$\bar{\cal{B}}^\prime \sim \bar{\cal{B}}$, and the baryon, $\cal{B}$.
The resulting superpotential in terms of these fields is then given 
to be :
\beq
\label{wduallow}
W = - {\cal{A}} \left( {\cal{B}} ~ \bar{\cal{B}}^\prime +
\bar{\Lambda}_1^{2 N} \right)
+  {\rm C}  \left(\bar{\Lambda}_2^{3M-N}
\cal{B} \right)^{1 \over M  } ~,
\eeq
where  C above is a constant.

We can now use the
constraint of eq. (\ref{wduallow}) to express $\cal{B}$ in terms of
$\bar{\cal{B}}^\prime$.  In addition, by  expressing $\bar{\cal{B}}^\prime$ in
terms of $\bar{\cal{B}}$, eq. (\ref{baryonmap}), and using  the scale matching
relations, eq. (\ref{sumscalematching}) and eq. (\ref{sunscalematching}), we
find that  the resulting superpotential is given (on appropriately identifying
the numerical constant C) by:
\beq
\label{gauginosun}
W_{gaugino} =M \left( {\Lambda_1^{2 N + M} \over \bar{\cal{B}}} \right)
^{1\over M}~.
\eeq
This superpotential has a natural explanation in terms of the
electric theory.  Since the $SU(N) $ group has $N-M$ flavors,
 we expect
a non-perturbative superpotential to arise due to gaugino condensation
in the unbroken
electric subgroup $SU(M) \subset SU(N)$.
This superpotential is exactly given by eq. (\ref{gauginosun})
\cite{seibergexact}.
Note that due to this superpotential, the quantum theory has
no moduli space.

\section{ Supersymmetry Breaking.}

 With this  understanding of the $SU(N) \times SU(M)$ dual
theory at hand, we  turn  in this section to the question of SUSY breaking.
We will mainly focus our attention on the $M=2$ case in which, 
as was discussed above, eq. (\ref{wtreebaryon}), all the flat directions can be raised
by adding appropriate terms in the tree level superpotential.
Among the $M=2$ theories we will analyze the odd-$N$ theories first,
then consider the even-$N$ theories.
Towards the end of this section we will briefly comment on the case
of general $M$ as well. 

\subsection{ The $M=2$   odd - $N$ Theories.}

We begin our analysis by returning to the superpotential eq. (\ref{wdual2}).
Recall that below the scale $\bar{\Lambda}_1$ the $SU(M=2)$ theory has 
$N$ flavors, $N_A$ and $\bar{r}^A$.  Furthermore,  as is clear from 
eq.~(\ref{wdual2}) all of them have mass and we expect the $SU(2)$ theory to be
driven into the  confining regime. In the subsequent discussion we will find it 
sometimes convenient  to adopt a common notation $U_i$ for all the quarks of the
$SU(2)$  group with $U_i=N_i /{\bar{\Lambda}_1}$ for $i\le N$ 
and $U_i=\bar{r}^{i-N}$ for $N<i \le 2N$. 
The mesons of $SU(2)$ will then 
be referred to as $V_{ij} \equiv U_i \cdot U_j $.  Since, as was 
mentioned above,  
one expects the $SU(2)$ theory to be driven into the confining regime
one can adequately account for the non-perturbative dynamics by working in
terms of the $SU(2)$ meson fields and adding a superpotential to the theory 
of the form:
\beq
\label{sanonpert}
W= \left( {{\rm Pf} V  ~ \bar{\Lambda}_1^N \over 
\bar{\Lambda}_{2}^{6-N} \cal{B}} \right)^{1\over N-2}.
\eeq
The $\cal{B}$ dependence above arises because the scale of the $SU(2)$ theory
is field dependent, eq. (\ref{lambda2l}).  The full superpotential 
is then given by a sum of eq. (\ref{wtreebaryon}),  eq. (\ref{wdual2}) and 
eq. (\ref{sanonpert}) to be:
\beqa
\label{safull}
W &=&~ N_A  \cdot \bar{r}^A +
{\cal{A}} \left( N^2 \cdot K^{N -2} - {\cal{B}} ~
\bar{\cal{B}}^\prime - \Lambda_{1 L}^
{2 N} \right) ~ \\
&&  + \left({{\rm Pf} V ~ \bar{\Lambda}_1^N \over \bar{\Lambda}_{2}^{6-N} 
                    \cal{B}}\right)^{1\over N-2} 
+  \mu ~ \lambda^{IA} ~ K_{AI} + \alpha_{AB} ~{\Lambda_2^{N-3} \over \mu}
 ~{\bar r}^A \cdot {\bar r}^B  
~\nonumber.
\eeqa
In the equation above we have set $Y_{IA} = \mu  K_{AI}$ and 
used the baryon operator map in SQCD,  \cite{seiberg}, \cite{dualityreview},
to write 
\beq
\label{antibaryonmap}
{\bar b}^{AB} = {\Lambda_2^{N-3} \over \mu} ~{\bar r}^A \cdot {\bar r}^B.
\eeq

We now restrict  our attention to the odd-$N$ case and show that the superpotential
in eq.~(\ref{safull}) implies that SUSY is broken.  We work in a basis in 
flavor space where $\lambda_{IA} = \lambda_I \delta_{IA}$ for $A < N-1$ 
and $\lambda_{IA} =0$ for $A \ge N-1$.  The matrix $\alpha_{AB}$ has rank 
$N-1$  with $\alpha^{AN}=0$. As was mentioned in the discussion following 
eq.~(\ref{wtreebaryon}) this choice of couplings lifts all flat directions in the 
classical theory.  Our strategy will be to   start by  assuming  that  all the $F$ term
conditions are valid  and   to solve for  some fields using them.  But doing so  
 will  lead to the conclusion   that the constraint enforced by 
 the Lagrange multiplier ${\cal{A}}$, namely 
 $N^2 \cdot K^{N -2} - {\cal{B}} ~\bar{\cal{B}}^\prime = \Lambda_{1 L}^{2 N}$,  
cannot be met  and therefore SUSY
must be  broken.

We begin by  noticing that the F term equation for $K_{A<NI}$ implies that 
${\cal{A}} \ne 0$.  From the equation for ${\bar {\cal{B}}}^{\prime}$ 
we learn then that  ${\cal{B}} =0$ and therefore that the second term in the 
constraint  vanishes.  We now prove that the first term vanishes too,
thereby showing that the constraint cannot be met.  For
this purpose notice first that  the equations for $K_{NI}$ imply  that  $(N^2 \cdot
K^{N-3})^{IN} =0$.  If in addition we can show that $N_A \cdot N_N$ vanishes
 then the first term in the constraint will have to vanish. 
In order to show  this, it is in fact convenient to consider 
the vevs of all the $SU(2)$ mesons  together. 
  Notice that the first term in eq. (\ref{safull}) is a  mass for the $N \cdot
{\bar r}$ mesons, the last term a mass for the $\bar{r} \cdot \bar{r}$ mesons, while
the second term in eq. (\ref{safull})
 can be regarded as  a mass  of order, ${\cal{A}}~ K^{N-2}$ for 
the $ N \cdot N $ mesons.  The vevs for the $SU(2)$ mesons can now be  expressed
in terms of these masses in the standard way.  On doing so, one finds that the 
expectation values of the $N_A \cdot N_N$ mesons  do indeed vanish.   This
completes the proof of SUSY breaking. 

A few  more points  need to be addressed with regards  to the above discussion.
 First, we did not allow for the possibility of a runaway vacuum, with some fields
going to  infinity,  in our discussion.  This should be a good assumption  since
we start with a theory in which classically all the flat directions  are  lifted.  Second, 
we assumed  that the K\"ahler potential  is  non-singular.  There  are in fact  some
points in moduli space where this  assumption is  invalid.  When 
$ {\cal{B}}\rightarrow 0$,
 eq. (\ref{lambda2S})  shows that    $\bar{\Lambda}_{2L}
\rightarrow 0 $  as well  and one expects a singularity in the K\" ahler potential since
extra fields will enter the low-energy theory.   We have analyzed these points
 in two  ways.  First, we added an extra flavor (with
a mass term)  for  $SU(N)$. In this case $\bar{\Lambda}_{2L}$  is not field
dependent and no singularity  arises in the K\" ahler potential  when 
${\cal{B}}\rightarrow 0$. Second,  since the scale of the $SU(2)$  group goes to
zero   at these points,  we worked  directly at the point ${\cal{B}}=0$ in terms of the
quarks of the $SU(2)$ theory. Both ways of analyzing the theory show that 
SUSY cannot be restored when ${\cal{B}}\rightarrow 0$.

  There is another  argument, involving an $R$ symmetry, which  shows that
SUSY must be broken in these theories.  As  was pointed out in the discussion following 
eq.~(\ref{wtreebaryon}), there is a flavor dependent $R$ symmetry that is left
unbroken in this case.  It turns out that all the fields involved in the constraint
implemented by the Lagrange multiplier $\cal{A}$ in eq. (\ref{safull}) are charged
under this $R$ symmetry. Thus  if this constraint is met the $R$ symmetry must be 
broken. In the absence of any flat directions one  concludes then that SUSY 
must be broken as well. The only alternative is that the constraint is not met,
but then again, SUSY must be broken.
The
behavior of these models is  therefore in accord with 
the considerations of ref. \cite{nelsonandseiberg}.

Finally, 
recall
that the $SU(N) \times SU(2)$ theory we have analyzed
here is  dual to  the  $SU(N)  \times SU(N-2)$  electric theory we started with.   The
low-energy degrees of  freedom and the superpotential identified in the $SU(N)
\times SU(2)$ theory should also provide a good description  of the infra-red
dynamics in the electric theory.  As was discussed 
in the beginning
 of section 3,  for
small enough  tree-level couplings, eq.~(\ref{wtreebaryon}),  the SUSY breaking
scale  is small too. 
The low-energy theory we considered is therefore a valid framework for
studying SUSY breaking, and
the electric theory  will  break SUSY as well. 

\subsection{ The $M=2$ even-$N$  Theories.}

We turn next to  the $M=2$ theories with   even $N$.  In this case, 
when all the flat directions are lifted there is no $R$ symmetry that is left 
unbroken.  
This might make one suspect that there is no SUSY breaking.
Indeed, by analyzing the  theory with the superpotential, eq.~(\ref{safull}), 
we can establish this result for a large class of 
couplings, $\lambda^{IA}$ and $\alpha_{AB}$, in eq.~(\ref{wtreebaryon})
(for which all the flat directions are lifted). 
While we do not present the details here, one finds 
 that all the $F$-term 
conditions can be met at a point where ${\cal{B}} \rightarrow  0$\footnote{
In terms of the description used above the Pfaffian of the mass matrix for 
the $SU(2)$ mesons goes to zero at this point  as well, thereby keeping the vevs of 
all the moduli fields finite.}. 
  The class of couplings for which we have been 
able to establish this  can be described as follows. As in the discussion 
following eq.~(\ref{wtreebaryon}), let us go to a basis where $\lambda^{IA}$ 
is non-zero when $A \le N-2$. In this basis as long as $\alpha_{AB}$ is 
zero when $A\le N-2$ and $B >N-2$ one can show that SUSY is restored. 
We strongly suspect this to be true in general. 

One legitimate concern about this analysis might be that, 
as  was mentioned in our discussion of the odd-$N$ theories, at points where
${\cal{B}} \rightarrow 0 $, there is a singularity in the K\"ahler potential and  the
effective Lagrangian  used  in  the analysis  breaks
down.   As in the odd-$N$ case, in order to address this concern, 
we study the theory in two ways. 
First, we add an extra flavor (with a mass term)  for  $SU(N)$.
In this case $\bar{\Lambda}_{2L}$ is not field dependent and
no singularity  arises in the K\" ahler potential  when ${\cal {B }}\rightarrow 0 $.
Again, we find that all the $F$-term conditions can be met at a point
where ${\cal{B}} \rightarrow 0 $. In fact, we recover the vevs obtained 
 in the analysis  above without the extra $SU(N)$ flavor,  after 
relating
the strong-coupling scales 
in the two cases. 
Second,  since the strong coupling scale of the $SU(2)$ theory 
tends
to zero 
when ${\cal{B} }\rightarrow 0 $, we
work directly at the point ${\cal{B}}=0 $, in terms of the quarks of 
$SU(M=2)$,  and verify  again that 
all the $F$-term constraints can be satisfied.

\subsection{The $M>2$ Theories. }

We end this section with some comments on the $M>2$ case. 
As was mentioned earlier, 
finding a tree-level superpotential that lifts
all the  baryonic flat directions 
 is considerably more  complicated  in this 
case,  and we have not been able to solve this problem yet.  One can  of course analyze
these theories with just the Yukawa couplings  eq. (\ref{wtree}).  However in this 
case there are baryonic flat directions and  a major concern is that these theories 
 have 
a  runaway  SUSY preserving vacuum\footnote{We are 
grateful to M. Dine and especially Y. Shirman for discussions in this regard.}. 
The situation here   is analogous to that for the $SU(N) \times SU(N-1)$ 
models   which  was discussed in some detail  in \cite{US}. Consequently,
we  discuss it only briefly here. 
   Let us return to the electric theory 
we started with in the presence of the tree level superpotential
eq. (\ref{wtree}).  The direction in question corresponds to taking  the 
$\bar{R}$ fields to  infinity  along a baryonic flat direction. 
The $SU(N-2)$ group is completely broken along this direction  while the 
$SU(N)$ group is strongly coupled with a scale that diverges asymptotically.  
As in \cite{US}
  one can satisfy all the $F$-term conditions along this 
direction  when the $\bar{R} $ fields go to  infinity. Moreover, 
a preliminary investigation suggests that  the corrections to the 
the classical K\"ahler potential for $\bar{R}$ are small along this direction 
leading to the conclusion that SUSY is probably  restored 
when $\bar{R} \rightarrow \infty$.

\section {The ${\bf SU(N) \times SU(N)}$ Models.}

For completeness, we also briefly discuss the $SU(N) \times SU(N)$ models.
In this case, one finds that SUSY is unbroken.
To see this, we note that the $SU(N) \times SU(N)$ models
are very similar to some of the $SU(2)\times SU(2)$
models---the ``$[1,1]$" models---first
considered in ref. \cite{ken}; see also \cite{US}. The moduli fields are the
mesons
$Y_{IA} = \bar{L}_I \cdot Q \cdot \bar{R}_A$, and the baryons, 
$X = {\rm det} Q$,   $\bar{B}_L = {\rm det} \bar{L}$ and  $\bar{B}_R = {\rm det}
 \bar{R}$.
The superpotential, which  can be derived as in  \cite{ken}, \cite{US}, is:
\beq
\label{suntimessun}
W = A \left( {\rm det} Y - \bar{B}_L ~ \bar{B}_R ~ X  +
 \bar{B}_L ~ \Lambda_R^{2 N}   + \bar{B}_R ~  \Lambda_L^{2 N}  \right)~,
\eeq
with $\Lambda_{L}$  ($\Lambda_{R}$) the scale of the first (second) group
respectively (defined
such that the coefficients of the last two terms in (\ref{suntimessun})
are unity), and $A$ is a Lagrange multiplier.
Upon perturbing (\ref{suntimessun}) with a maximal rank Yukawa coupling,
$\delta W = \lambda^{IA} Y_{IA}$, one finds that there is a SUSY preserving
runaway vacuum along $X \rightarrow \infty$.
This runaway direction corresponds to a flat direction present in the
classical theory.
One could try to
stabilize the $X$ runaway
direction, by adding another term, proportional to $X$, to the
superpotential. But in this
case the theory has a SUSY preserving vacuum without any runaway behavior.

\section{Conclusions.}

In this paper,  we have studied a large class of $N=1$ supersymmetric theories
with gauge group  $SU(N) \times SU(N - M)$
and fundamental matter content. We used duality to 
elucidate  their nonperturbative dynamics. 
We showed that the odd-$N$, $M=2$
 theories  break supersymmetry after adding
appropriate Yukawa couplings. 
We expect that a similar analysis is applicable in other classes of product-group 
theories as well, e.g. the 
$SU(N) \times SP(M)$ theories \cite{it}. 
Our results provide more examples of theories that 
break supersymmetry and suggest that this phenomenon is fairly common in
chiral supersymmetric gauge theories. 
We hope that these examples will prove of use
in the construction of phenomenologically relevant models, 
and will  contribute to a systematic understanding 
 of supersymmetry breaking.

We would like to thank M. Dine and Y. Shirman for insightful 
discussions. 
E.P. acknowledges support by a Robert R. McCormick
Fellowship and by DOE contract DF-FGP2-90ER40560. 
Y.S. and S.T. acknowledge
the support of DOE contract DE-AC02-76CH0300.

\nc{\ib}[3]{ {\em ibid. }{\bf #1} (19#2) #3}
\nc{\np}[3]{ {\em Nucl.\ Phys. }{\bf #1} (19#2) #3}
\nc{\pl}[3]{ {\em Phys.\ Lett. }{\bf #1} (19#2) #3}
\nc{\pr}[3]{ {\em Phys.\ Rev. }{\bf #1} (19#2) #3}
\nc{\prep}[3]{ {\em Phys.\ Rep. }{\bf #1} (19#2) #3}
\nc{\prl}[3]{ {\em Phys.\ Rev.\ Lett. }{\bf #1} (19#2) #3}

\end{document}